\documentclass[conference]{IEEEtran}

\usepackage{cite}
\usepackage{amsmath,amssymb,amsfonts}
\usepackage[ruled,vlined]{algorithm2e}
\usepackage{graphicx}
\usepackage{textcomp}
\usepackage{xcolor}
\usepackage{times}
\usepackage{soul}
\usepackage{url}
\usepackage[utf8]{inputenc}
\usepackage[small]{caption}
\usepackage{pifont}
\usepackage{booktabs}
\usepackage{mathtools}
\usepackage[noend]{algpseudocode}
\usepackage{tikz}
\let\norm\undefined 
\DeclarePairedDelimiter\norm{\lVert}{\rVert}
\newcommand{\argmin}{\mathop{\mathrm{argmin}}\limits}
\def\BibTeX{{\rm B\kern-.05em{\sc i\kern-.025em b}\kern-.08em
    T\kern-.1667em\lower.7ex\hbox{E}\kern-.125emX}}
    
\newcommand\copyrighttext{%
  \footnotesize \textcopyright 2020 IEEE. Personal use of this material is permitted.
  Permission from IEEE must be obtained for all other uses, in any current or future
  media, including reprinting/republishing this material for advertising or promotional
  purposes, creating new collective works, for resale or redistribution to servers or
  lists, or reuse of any copyrighted component of this work in other works.}
\newcommand\copyrightnotice{%
\begin{tikzpicture}[remember picture,overlay]
\node[anchor=south,yshift=10pt] at (current page.south) {\fbox{\parbox{\dimexpr\textwidth-\fboxsep-\fboxrule\relax}{\copyrighttext}}};
\end{tikzpicture}%
}
    
\begin{document}

\title{Distributed Machine Learning for Predictive Analytics in Mobile Edge Computing Based IoT Environments}

\author{\IEEEauthorblockN{Prabath Abeysekara}
\IEEEauthorblockA{\textit{School of Science} \\
\textit{RMIT University}\\
Melbourne, Australia \\
s3693452@student.rmit.edu.au}
\and
\IEEEauthorblockN{Hai Dong}
\IEEEauthorblockA{\textit{School of Science} \\
\textit{RMIT University}\\
Melbourne, Australia \\
hai.dong@rmit.edu.au}
\and
\IEEEauthorblockN{A. K. Qin}
\IEEEauthorblockA{\textit{Department of Computer Science and Software Engineering} \\
\textit{Swinburne University of Technology}\\
Melbourne, Australia \\
kqin@swin.edu.au}
}

\maketitle
\copyrightnotice

\begin{abstract}
Predictive analytics in Mobile Edge Computing (MEC) based Internet of Things (IoT) is becoming a high demand in many real-world applications. A prediction problem in an MEC-based IoT environment typically corresponds to a collection of tasks with each task solved in a specific MEC environment based on the data accumulated locally, which can be regarded as a Multi-task Learning (MTL) problem. However, the heterogeneity of the data (non-IIDness) accumulated across different MEC environments challenges the application of general MTL techniques in such a setting. Federated MTL (FMTL) has recently emerged as an attempt to address this issue. Besides FMTL, there exists another powerful but under-exploited distributed machine learning technique, called Network Lasso (NL), which is inherently related to FMTL but has its own unique features. In this paper, we made an in-depth evaluation and comparison of these two techniques on three distinct IoT datasets representing real-world application scenarios. Experimental results revealed that NL outperformed FMTL in MEC-based IoT environments in terms of both accuracy and computational efficiency.
\end{abstract}


\section{Introduction}
\label{introduction}
\begin{figure}[hbt!]
    \centering
    \includegraphics[height=0.25\textwidth]{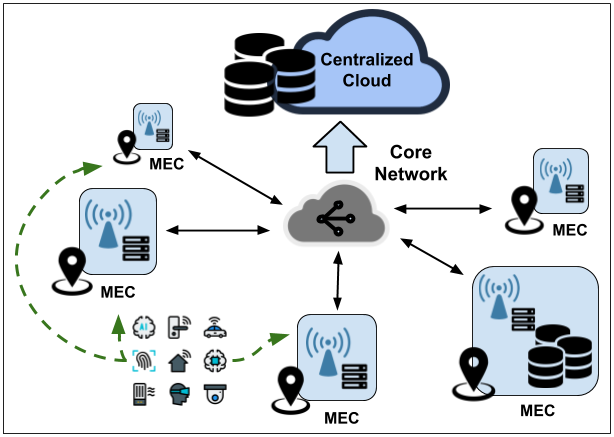}
    \caption{The typical structure of a MEC topology.}\label{fig2}
\end{figure}

Mobile Edge Computing (MEC) based Internet of Things (IoT) is a rapidly developing paradigm to providing geographically distributed computing and storage resources to smart devices and applications at the \textit{edge} of the network \cite{RN23}. By allowing high-volume data generated by a massive amount of networked smart devices (e.g., mobile devices, self-driving cars and fitness trackers) in modern IoT systems to be processed at the edge, it promises to greatly reduce the network stress on the core networks of mobile service providers. Acting as the first line of defense for high-volume IoT data, MEC infrastructures allow performing computationally intensive data processing tasks such as data filtering, as well as persistence of raw data at the \textit{edge} and, at times, transmit significantly lesser amounts of data at lower frequencies to the cloud environments for further analytics. This data accumulated at the edge of the network inherently creates a \textit{distributed} topology of data silos across different MEC environments (see Fig. \ref{fig2}).

In modern IoT systems, machine learning based predictive analytics becomes popular to solve challenging problems and uncover new patterns for people and enterprises alike. However, as to the use of MEC in modern IoT systems, the \textit{distributedness} of data sources it introduces and complex communication topologies with which these MEC environments are linked to each other make existing predictive analytics strategies obsolete in such a context \cite{RN323}. For example, most traditional machine learning strategies designed for existing cloud-based IoT systems demand the data to be accumulated centrally. This undisputedly defies the goals of MEC as it causes significant stress on the core mobile networks due to the data generated in large volumes by the IoT devices. This calls for efficient and scalable machine learning strategies that not only fit into a distributed setting, but also can perform atop high-volume data.

In addition, \textit{most machine learning strategies proposed for IoT systems assume that the data generated by sensor service providers and consumers across the entire network can be represented by a single distribution}. As a result, they often hypothesize that a single global prediction model can represent the characteristics of the underlying application domain (eg. a global trust prediction model used by self-driving cars to predict trustworthiness of all IoT services). This one-size-always-fits-all approach based on independent and identically distributed (IID) assumptions of the data is often restrictive in MEC-based IoT systems \cite{RN319}. For instance, the heterogeneity of sensor service providers and consumers taking part in transactions can potentially result in datasets with distinct characteristics (i.e. non-IID) in different MEC-environments. Hence, there is an apparent need for strategies that allow training high-quality \textit{decentralized local models} that fit into the nature and structure of each distributed datasets accumulated across different MEC environments. 

Furthermore, \textit{some MEC environments can accumulate data from sensor service providers and consumers with similar configurations and characteristics, thereby producing similarly-poised prediction models}. In such a context, letting similar models exchange strength with each other has the potential of improving the generalization performance of them. In our work, we formally define the notion of \textit{exchanging strength} as the act of sharing models trained by individual MEC environments directly with peers (eg. neighboring MEC environments) or an intermediary (eg. centralized cloud) over complex communication topologies. We postulate that the exchanged strength (i.e. knowledge) can be suitably aggregated or adapted as is by the receivers to train prediction models with better accuracy for each individual MEC environment. This demands for machine learning strategies that encourage knowledge sharing among distributed prediction models trained across such networked systems with complex communication topologies.

Lies in the intersection of achieving a good balance between the aforementioned verticals is Federated Multi-task Learning (FMTL). Combining the power of Federated and Multi-task Learning paradigms, FMTL attempts to collaboratively learn high-quality distributed prediction models that best represent the data that they are trained upon \cite{RN319}. FMTL is built on the principle idea of decomposing a finite-sum problem defined over a set of distributed agents, and letting each one of those agents independently solve a smaller sub-problem over the data accumulated within its own context. The hope is such a mechanism would allow these distributed agents (i.e. workers) to process its own data closer to where it originated and share only a fraction of data with a centralized (i.e. master) server for further processing in a communication-efficient manner. In addition, FMTL also allows either fixing or learning relationships among these distributed agents thereby allowing them to share knowledge with similar others and being able to produce prediction models with better quality. As a result, FMTL can be deemed a befitting paradigm to address the challenges posed by MEC systems.

Besides FMTL, there exists another less popular but powerful distributed machine learning technique, called Network Lasso (NL), which is inherently related to FMTL but has its own unique features. Originally proposed to achieve simultaneous clustering and optimization in large-scale networks of distributed agents, in combination of the well-known Alternating Method of Multipliers (ADMM), it provides a solid algorithmic framework to collaboratively derive a family of machine learning models over a distributed set of agents, in parallel \cite{RN192}. In addition, it also boasts global convergence of any arbitrary problem that can be represented in its algorithmic framework. Therefore, \textit{it can be introduced as a seemingly elegant framework that can address the challenges in using machine learning in MEC-based IoT environments, which has not been studied before in the aforementioned context}. Therefore, in this work, we 
\begin{itemize}
    \item analysed and compared these two techniques from a methdological and algorithmic perspective, and 
    \item performed an in-depth empirical study on IoT datasets representing three distinct real-world application scenarios to reveal the performance difference between these two techniques. 
\end{itemize}
The analysis and comparison results demonstrate that, although, Mocha FMTL framework promises enticing features that can greatly enhance its suitability to an MEC-based IoT environment, the higher accuracy and ability to achieve a better prediction performance in comparatively fewer communication iterations make NL a more suitable choice for the aforementioned setting. 

The remainder of this paper is organized as follows. 
Section \ref{technical-preliminaries} provides a brief technical summary of the two approaches, before discussing the same extensively at a system- and algorithmic framework-level in Section \ref{methdology-comparison} and \ref{algorithm-comparison}, respectively. Section \ref{evaluation} introduces the framework used for the empirical evaluation of the two approaches and Section \ref{results-and-discussion} discusses the results gathered. Section \ref{conclusion} concludes our work.

\section{Technical Preliminaries}
\label{technical-preliminaries}
Before we delve further into the comparison between FMTL and NL, we provide a brief systematic exposition below on them and the fundamental principles atop which they are built, for completeness.

\subsection{Multi-task Learning}
Multi-task learning (MTL) paradigm is built on the principle of allowing multiple different tasks to collaborate with themselves to improve the generalization performance of their respective models \cite{RN326}. Such a collaboration allows these tasks to train a single shared prediction model, or multiple related models in the context of clustered multi-task learning, together \cite{RN325}. In other words, it is expected that some relationships exists among all or some of these tasks, which can be exploited in a favorable way that each participating task benefits from it. For instance, the data generation mechanism of the tasks can be approximately similar, which, in the context of machine learning, is usually considered a fair metric to assume that the phenomenon each task is representing is similar, as well and vice versa. Therefore, similarly-poised tasks can let each other exchange strength to train prediction models with better performance. On the other hand, dissimilar tasks can also repel each other and attempt to be clustered with those that are similar to them.

Given many variants proposed in the current literature to achieve Multi-task Learning, a unified formulation for regularized Multi-task Learning can be defined, as below \cite{RN319, RN326}.

\begin{equation}\label{mtl}
\begin{array}{c}
\underset{W,\Omega}{\text{min}} \displaystyle\bigg\{\sum\limits_{t=1}^{m} \ell_{t}(w_{t}x_{t}, y_{t}) + R(W, \Omega)\bigg\}.
\end{array}
\end{equation}
where, $t\in\{1,\dots,m\}$ denotes the tasks taking part in prediction model training, $\{x_{t}^{i},y_{t}^{i}\}_{i=1}^{n_t}$ denotes the training dataset of $t^{th}$ task in that $x_t \in \mathbb{R}^{d}$ and $y \in \mathbb{R}$, $\ell_t$ is a convex loss function minimized by the $t^{th}$ task such as, hinge-loss in Support Vector Machines (SVM), $w_t$ represents the parameters of the model trained by the $t^{th}$ task, $W \in \mathbb{R}^{d \times m}:=[w_1,w_2,\dots,w_m]$ is a matrix representing the model parameters of all the tasks in which its $t^{th}$ column carries the model parameters of the $t^{th}$ task. $\Omega \in \mathbb{R}^{m \times m}$ denotes the pairwise relationships among the tasks in the form of a covariance matrix, which is either known before hand, or learnt during the learning process.

\subsection{Federated Multi-task Learning}

\begin{algorithm}
\SetAlgoLined
 \For{k = 1,\dots,K}{
   \For{t = 1,\dots,m}{
     $\Delta \alpha_t =$ solved used in each distributed node returns $\Theta_{t}^{k}$-approximate solution $\Delta \alpha_t$\\
     $\alpha_t = \alpha_t + \Delta \alpha_t$\\  
     return $\Delta v_t = X_t\Delta\alpha_t$
   }
   $v_t = v_t + \Delta v_t$\\
   Update $\Omega$ centrally based on $w(\alpha)$ for latest $\alpha$\\
 }
 Update $w = w(\alpha)$ based on latest $\alpha$\\
 Return $W = \{w_1, w_2, \dots, w_m\}$
 \caption{Mocha - FMTL Framework}
 \label{alg:mocha}
\end{algorithm}

Built on the same unified Multi-task Learning formulation in \eqref{mtl}, FMTL could be looked at as an approach to address key 1) statistical challenges such as non-IID (non Identically and Independently Distributed) and unbalanced data as well as 2) systems challenges such as unreliable network connectivity and resource constraints of multiple agents (i.e. networked IoT devices, mobile phones, etc) in a federated setting to enable Multi-task Learning more efficiently \cite{RN319}. The first work that formally introduced the FMTL paradigm proposed the algorithmic framework named Mocha (see Algorithm 1) \cite{RN319} to address the proclaimed statistical and systems challenges in a federated setting.

The primary focus of Mocha FMTL framework is to address Multi-task Learning problems in which the task relationships denoted by $\Omega$ in \eqref{mtl} are learnt dynamically. However, it can also be applied to problems where the relationships are static and pre-configured, as well. In such a context, the problem formulation \eqref{mtl} could be deduced to,
\begin{equation}\label{static-fmtl}
\begin{array}{c}
\underset{W}{\text{min}} \displaystyle\bigg\{\sum\limits_{t=1}^{m} \ell_{t}(w_{t}x_{t}^{i}, y_{t}^{i}) + R(W, \Omega)\bigg\}.
\end{array}
\end{equation}
where, $\Omega$ is no longer learnt simultaneously with the model parameters $w_{t}$ of the tasks, and therefore, is static.

\subsection{Network Lasso}

Network lasso is a framework to solve large-scale optimization problems formulated as a graph structure, allowing simultaneous clustering and optimization \cite{RN192}. Given an un-directed networked graph, i.e. $\left. G=(V,E)\right.$ in which nodes are denoted by i.e. $\left. V=\left\{1,\dots,N\right\}\right.$, and their connectivity with each other is denoted by the edges i.e. $\left. E=\left\{(v_1, v_2): v_1,v_2 \in V, v_1 \neq v_2\right\}\right.$, the NL problem is mathematically expressed, as below.

\begin{equation}\label{network-lasso}
\begin{array}{ccclcl}
{\text{minimize}} & \displaystyle\sum\limits_{t \in V} \ell_{t}(w_{t}) + \lambda\sum\limits_{(j,k) \in E}  a_{jk}||w_{j} - w_{k}||_{2}.
\end{array}
\end{equation}

In this optimization problem, $\left. w_t \in \mathbb{R}^n\right.$ and represents model parameters of a convex loss function $\ell_t$. Each loss function $\left.\ell_{t}\right.$ defined over the input-output space $\left. \ell_t: \mathbb{R}^n \to \mathbb{R} \cup \left\{\infty\right\} \right.$ is local to a node $\left. v_{t} \in V \right.$  in the graph $\left. G \right.$. These loss functions are used to estimate model parameters of each node in the graph $\left. v_{t} \in V \right.$ by formulating an optimization problem.  Meanwhile, $\left.\lambda\right.$ is a regularization parameter that scales the edge objectives relative to the node objectives, $\left. a_{jk} \right.$ represents an impact factor of a particular edge i.e. $\left. (v_j,v_k) \right.$ on the finite-sum problem computed over the loss functions of all nodes participating in the optimization problem. It is also noteworthy that $\left.w_j,w_k\right.$ correspond to the parameters of the models associated with two adjacent nodes $\left.v_j,v_k\right.$ in the graph, respectively. Furthermore, the regularization parameter $\left.\lambda\right.$, impact factor $\left. a_{jk} \right.$ alongside the $\left.\ell_{2}\right.$-norm computed over the difference of model parameters between the two nodes connected by the edge $\left. (v_j,v_k) \right.$ form a penalty factor. This compels the contrast between two connected nodes to be zero strengthening the cohesion among those that carry similar model parameters (i.e. $\left.w_j=w_k\right.$). 

\begin{algorithm}
\label{alg:network-lasso-admm}
\SetAlgoLined
 \While{$\norm{r_{p}^{k}}_{2} < \epsilon^{p}$ and $\norm{r_{s}^{k}}_{2} < \epsilon^{d}$}{
  \begin{align}
w_{t}^{k+1} &= \argmin_{w_{t}} \Big\{\ell_{t}(w_{t})+ \sum\limits_{i \in N(t)} \frac{\rho}{2}\norm{w_{t} - z_{ti}^{k} + u_{ti}^{k}}_{2}^{2}\Big\}\nonumber\\
z_{ti}^{k+1} &= \theta(w_{t} + u_{ti}) + (1-\theta)(w_i + u_{ti})\nonumber\\
z_{it}^{k+1} &= (1-\theta)(w_{t} + u_{ti}) + \theta(w_i + u_{ti})\nonumber\\
u_{ti}^{k+1} &= u_{ti}^{k} + (w_{t}^{k+1} - z_{ti}^{k+1})\nonumber
\end{align}
 }
 \caption{Network Lasso parallelized by ADMM}
\end{algorithm}

By default, NL uses ADMM \cite{RN211} to decompose problem \eqref{mtl} into smaller sub-problems in which each task solves its own data-local sub-problem in parallel, passes the solution to its neighbors, and repeats the process until the entire network converges. This gives rise to the parallel algorithm expressed in Algorithm 2 \cite{RN192}, in which $r_{p}^{k}$ and $r_{s}^{k}$ correspond to the primal and dual residuals, which are typically used to determine the stopping criteria of the algorithm \cite{RN211}. A suitable stopping criteria allows achieving a favorable trade-off between accuracy and number of iterations, which directly contributes to communication cost in networked systems.

The \textit{network lasso penalty} (i.e. $||w_{j} - w_{k}||_{2}$) defined over edges representing neighboring nodes in problem \eqref{network-lasso} makes it all the more interesting, as it encourages them to be similar \cite{RN192}. This allows a node to \textit{exchange strength} with a suitable neighbor, in the form of consensus. Eventually, the neighboring nodes that are in consensus with each other (i.e. nodes with similar data distributions) will form clusters together while those that aren’t (i.e. nodes with different data distributions) will form different clusters. 

\section{Methodology Comparison}
\label{methdology-comparison}
In this section, we intend to compare Mocha FMTL framework and NL framework against their suitability to the key characteristics of MEC-based IoT environments announced in Section \ref{introduction} at a solution level. 

\noindent\textbf{Knowledge sharing under non-IID data:}
Both Mocha and NL, by default, allow knowledge sharing among tasks, even though the semantics of which are somewhat different. For instance, Mocha attempts to perform knowledge sharing via aggregating the models of similarly-poised tasks, which are either configured statically or learnt dynamically. In contrast, NL uses Simultaneous Clustering and Optimization (SCO) for the same purpose \cite{RN192}. In other words, each task in a given statically formed task topology attempts to adapt a suitable model (determined through comparing the geometric similarity of its own model against that of its neighbours' models) of a neighbouring task, and use that as its own model. As a result, similarly-poised tasks (i.e. tasks that adapted the same model from a neighbor) will form clusters while those that are different will form suitable other clusters. Therefore, it could be deduced that both approaches are capable of handling non-IID data available in MEC-based IoT environments, suitably. However, the fact that Mocha can also learn the structures of similarly-poised tasks and enable knowledge sharing dynamically, as opposed to the static approach followed by NL gives it the edge.


\begin{figure}[h]
    \centering
    \includegraphics[height=0.24\textwidth]{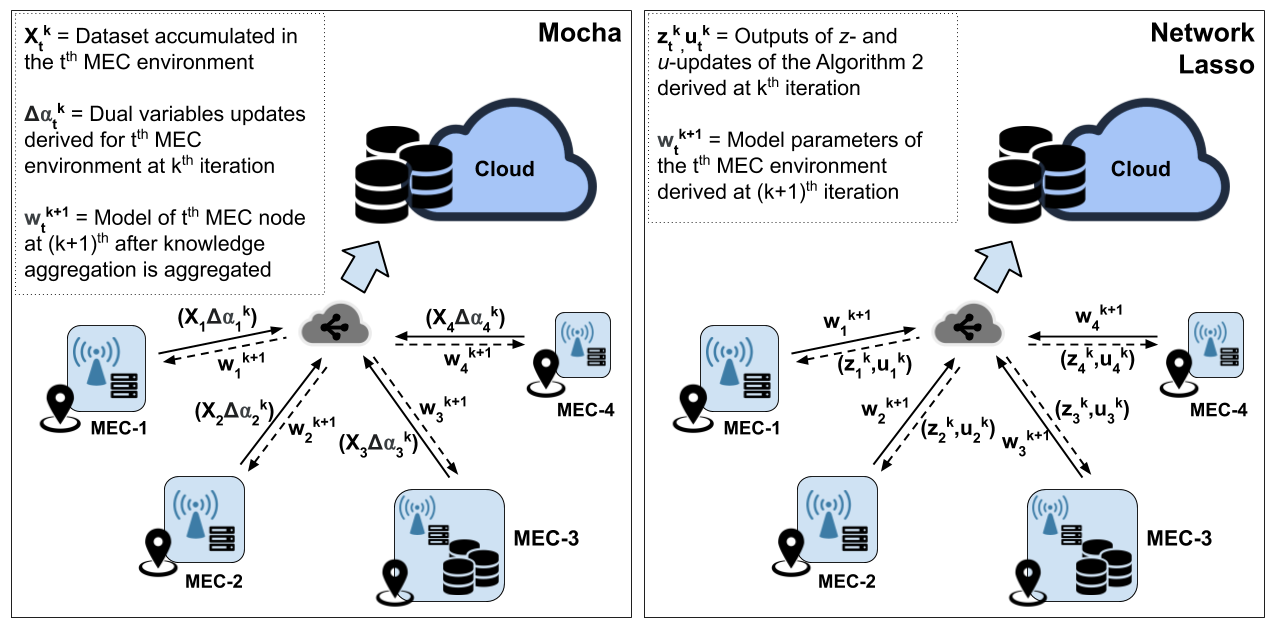}
    \caption{The information flow in Mocha FMTL and NL frameworks over a hierarchical network topology}\label{fig9}
\end{figure}

\noindent\textbf{Support for hierarchical network topologies:}
Mocha FMTL framework, by design, is built on top of a \textit{client-server} architecture. In that, the primary responsibility of the tasks (i.e. clients) are to iteratively train a suitable prediction model atop the locally accumulated data and share the model parameters with a central server for knowledge sharing. Then, the central server facilitates knowledge sharing among similarly-poised tasks (the topology of which is either statically linked or learnt dynamically) as well as learning relationships among different tasks. Therefore, Mocha can be deemed adequately suitable to the network hierarchy typically used by MEC-based IoT environments, which is depicted in Fig-\ref{fig2}.

In contrast, NL framework, by default, encourages \textit{direct communication among the neighbouring tasks} (i.e. clients) for knowledge sharing, with minimal assistance from a centralized server only to coordinate the iterations of the algorithmic framework shown in Algorithm 2 \cite{RN192}. Due to the challenges described in Section \ref{introduction}, allowing direct communication among different MEC environments in an MEC-based IoT system can be tedious. However, \cite{RN323} proposes a hierarchical architecture for NL framework involving MEC environments and a centralized cloud layer in which, w-update of the Algorithm-2 is run in a distributed manner while z- and u-updates are run in the centralized cloud layer (see \ref{fig9}. This way, NL framework too can be transformed into a client-server like model described in Section \ref{introduction}. Therefore, it could be deduced that both frameworks possess the potential of being used in the concerned problem context. 

\noindent\textbf{Fault tolerance and straggler avoidance:}
MEC systems often involve complex networking infrastructure that are prone to failures (i.e. base stations being offline, etc.) \cite{RN328}. In addition, due to the heterogeneous resources and networking infrastructure used by different MEC environments, the quality of services (QoS) levels associated with the services exposed by them can vary considerably \cite{RN329}. Such factors can either lead to abrupt failures in synchronous algorithms or introduce stragglers thereby making it challenging to run distributed machine learning frameworks in MEC-based IoT systems efficiently. In this particular context, Mocha FMTL framework offers built-in support for straggler avoidance, which is a favorable factor for a typical MEC-based IoT environment.

In contrast, NL parallelized by ADMM, in its default form, is a synchronous algorithm, which is prone to catastrophic failures in case of a task failure. Therefore, Mocha seems a better choice for the concerned setting than the default NL framework. However, there is an opportunity to use asynchronous variants of ADMM to parallelize NL framework \cite{RN335}, which would improve the adaptability of it in the context of MEC-based IoT systems.

\section{Algorithm Comparison}
\label{algorithm-comparison}
This section reviews some prominent features in the algorithmic frameworks used by Mocha FMTL and NL. For simplicity, we refer to any arbitrary Multi-task Learning formulation parallelized by Mocha FMTL framework as \textit{Mocha}, and Network Lasso Parallelized by ADMM as \textit{NL}.

\noindent\textbf{Use of duality:} Both Mocha and NL are designed based on the principle of duality. However, there exists a key difference between the two approaches in the way the duality is utilized within their respective algorithmic frameworks. Mocha primarily employs the dual form of a given optimization problem (eg. regularized hinge-loss) and solves the resulting dual problem to arrive at an optimal solution using a variant of dual coordinate ascent optimization method. In contrast, NL parallelized by ADMM inherits the properties of the dual-subgradient based dual ascent method and can be viewed as a primal-dual optimization method. Therefore, it can be useful even when the dual of a given primal problem in closed form cannot be derived, and also in some cases where the dual problem is not differentiable \cite{RN211}.

\noindent\textbf{Problem decomposition strategy:} In a typical MEC-based IoT environment, every MEC environment accumulates data from IoT devices and stores them within MEC-local data centers. As a result, one key systems challenge that distributed machine learning algorithms for MEC-based IoT systems (or any other distributed system) need to cope with is efficiently training prediction models atop datasets accumulated in a decentralized manner. Therefore, it is essential to analyze how Mocha and NL achieve the aforementioned goal.

NL inherits its decomposability properties from dual decomposition, a vital precursor upon which ADMM is built \cite{RN211}. Together with variable splitting, ADMM derives a sound mathematical framework to decompose the NL problem into three key sub-problems denoted as w-, z- and u-updates in Algorithm 2 \cite{RN192}. In a typical MEC topology, $w$-update can be carried out within each distributed MEC environment in parallel, while $z$- and $u$-updates can be run in the centralized cloud layer. Mocha, on the other hand, uses a \textit{quadratic approximation} of the dual of the general Multi-task Learning framework denoted by problem \eqref{mtl}, as below. Let the dual of problem \eqref{mtl} be

\begin{equation}\label{dual-mtl}
\begin{array}{c}
D(\alpha) = \underset{\alpha}{\text{min}} \bigg\{\displaystyle\sum\limits_{t=1}^{m}\sum\limits_{i=1}^{n_t} \ell_{t}^{*}(-\alpha_t^i)+ R^*(X\alpha)\bigg\}.
\end{array}
\end{equation} 

where $\ell_t^*$ and $R*$ are (Fenchel) conjugate dual functions of $\ell_t$ and $R$, respectively. $\alpha_i$ corresponds to the dual variable associated with the training sample $(x_i, y_i)$. $X \in R^{mdxn}$ is defined as $X := diag(X_1, X_2,.., X_m)$. The number of training samples available in $t^{th}$ task is defined as $n_t$. To come up with distributed task-local sub-problems, a quadratic approximation of \eqref{dual-mtl} is then derived as below \cite{RN311}. 

\begin{equation}\label{w-update-mocha}
\begin{array}{c}
\Delta\alpha_{t} = \underset{\Delta\alpha_t}{\text{argmin}} \bigg\{\displaystyle\sum\limits_{i=1}^{n_t} \ell_{t}^{*}(-\alpha_t^{i} -\Delta\alpha_{t}^{i})+\big<w_t(\alpha),X_t\Delta\alpha_{t}\big> \\ + \frac{\sigma^{'}}{2}||X_t\Delta\alpha_t||_{M_t}^{2} + \frac{1}{m}R^*(X\alpha)\bigg\}.
\end{array}
\end{equation} 

In \eqref{w-update-mocha}, each and every component can be solved using the task-local information. Even though solving $\frac{1}{m}R^*(X\alpha)$ involves access to the $X\alpha$ which is defined over all the training examples across the tasks, the value of it computed in the previous step and shared by the centralized layer is used.

\noindent\textbf{Alternating minimization of sub-problems:} 
Both Mocha and NL follow an alternating minimization approach to solve problem \eqref{mtl}. However, there is a subtle difference between the exact method in which this happens in each of the aforementioned frameworks. For instance, NL framework solves three sub-problems $w$-, $z$- and $u$-updates derived by ADMM via proximal steps, in an alternating manner iteratively \cite{RN192}. While doing so, the solution of the preceding step is fixed and passed onto the next step until the next iteration kicks in. In this procedure, the task of knowledge sharing is baked into the optimization routines executed as part of the aforementioned sub-problems. 
In contrast to NL, Mocha uses only one sub-problem derived via proximal splitting. In addition, Mocha also decouples the task of solving the distributed sub-problem from the process of knowledge sharing. In that, it first solves the distributed sub-problems denoted by \eqref{w-update-mocha} and shares the solution with the centralized model aggregation layer similar to that of NL. Then, in contrast to NL, it fixes the shared model updates and attempts to learn the relationships among the tasks as a non-optimization step to enforce knowledge sharing. 

\noindent\textbf{Exact vs approximate derivation of sub-problems:} 
The sub-problems associated with NL have been formally derived by alternatingly minimizing the associated augmented Lagrangian of the NL problem \cite{RN192} atop a sound mathematical framework. This guarantees the \textit{exactness} of the sub-problem optimization as well as the ability of the algorithm to eventually arrive at a provably optimal stationary solution. However, it may come at a cost as the exact algorithms can at times suffer from comparatively slower convergence speeds against their inexact variants. On the other hand, the only sub-problem used by Mocha is derived by a suitable quadratic distributed approximation of \eqref{dual-mtl}. Consequently, the hope is that the optimization of the approximated sub-problem may eventually arrive at a stationary solution with reasonable guarantees on the accuracy and  convergence speed. Therefore, from a MEC perspective, the empirical performance of the two approaches can be used as a metric to decide the best suited framework for the above outlined setting.

\noindent\textbf{Solvers for data-local sub-problems:} 
Mocha, in its default implementation, relies on Stochastic Distributed Coordinate Ascent (SDCA) \cite{RN337} as the optimization method used to solve the task-local sub-problems. In contrast, NL parallelized by ADMM uses standard Gradient Descent (GD), and therefore, it primarily can be labelled as a batch method. At each iteration of the algorithm, it computes a distributed batch gradient, which can be computationally expensive. This can be significantly restrictive in the context of large-scale settings such as MEC-based IoT systems where data is accumulated in higher volumes at rapid velocities. However, many stochastic variants of ADMM had been introduced in the prior literature in order to reduce the computational complexity of ADMM \cite{RN270, RN298}. However, their ability to ensure communication efficiency in environments such as MEC-based IoT systems needs to be further investigated.

\section{Evaluation}

\begin{figure*}
 \includegraphics[width=1\linewidth,keepaspectratio]{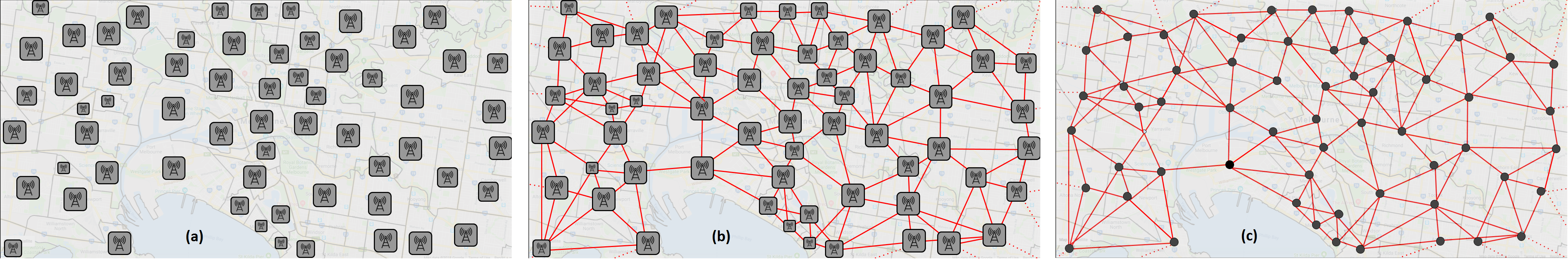}
 \caption{A hypothetical deployment of MEC environment, which shows how the neighbouring MEC environments are linked based on proximity forming a partial-mesh network.}\label{fig1}
\end{figure*}

\label{evaluation}
We conducted a series of experiments to evaluate the suitability of Mocha FMTL and NL frameworks for MEC-based IoT environments. The primary objective of these experiments was comparing the performance of each framework against two key systems and statistical challenges of using distributed machine learning in MEC-based IoT environments. We primarily evaluated the following key requirements of such systems identified in Section \ref{introduction}.
\begin{itemize}
\item \textbf{Support for distributed and Non-IID data}, which is a reflection of the different data generation mechanisms in distributed MEC-based IoT environments caused by the heterogeneity of IoT sensor service providers and consumers connecting to them.
\item \textbf{Communication efficiency}, which looks at the number of communication iterations corresponding to the communication between the centralized cloud and distributed MEC layers till convergence, and therefore, is an indicator of network stress on the core mobile networks of mobile network providers.
\end{itemize}

\subsection{Experiments}
We designed the following experiments to reflect the key characteristics highlighted before. Each experiment measures and compares the performance of Mocha FMTL and NL in a simulated MEC-based IoT environment. 

\begin{itemize}
\item To test the performance in the presence of Non-IID data, 100 distributed binary SVMs representing a distributed prediction models, were trained using both Mocha FMTL and NL frameworks atop the same splits. Prediction accuracy was used as the Key Performance Indicator (KPI) to compare the performance of each binary SVM classifier trained. It is important to note that both Mocha FMTL and NL frameworks were run with the recommended default hyper-parameter values outlined in their respective original works \cite{RN319, RN192} where relevant.
\item To test the communication efficiency, we kept track of the number of iterations needed by each framework until convergence, which is the primary KPI for the communication efficiency comparison between the two approaches.
\end{itemize}

Apart from the two main frameworks being compared, we also evaluated the two following baselines.

\begin{itemize}
\item 100 distributed binary SVM classifier trained atop the same splits of dataset upon which Mocha FMTL and NL models were trained, representing a distributed yet non-communicative and isolated set of prediction models,
\item a global binary SVM classifier atop the same aggregate datasets to represent a centralized prediction model.
\end{itemize}
Prediction accuracy was again used as the primary KPI to compare the performance of each scenario above.

The simulation, meanwhile, took into account a hypothetical MEC topology deployed across 100 suburbs in the Melbourne City Council area (see Fig-\ref{fig1}(a)). In this MEC topology, we hypothesized that every suburb has one MEC environment each deployed in them. We then connected each MEC environment to their most nearest 5 neighbours based on proximity (see Fig-\ref{fig1}(b)). The resulting MEC topology then forms a complex graph resembling a partial mesh network in which every node is connected to 5 other nodes based on proximity (see Fig-\ref{fig1}(c)). It is worth noting that, a more pragmatic deployment of a connected MEC-network depends heavily on a multitude of factors such physical deployment constraints and other Key Performance Indicators (i.e. latency, etc.) \cite{RN226}. However, we omitted such details for brevity and clarity as those deployment characteristics do not have any direct impact on information processing related tasks.

We used an experimental set-up that trained a binary SVM classifier at every MEC environment in the above MEC topology. The set-up primarily consisted of an application implemented in Python using CVXPY, a well-known Python-based modelling language and framework for solving convex optimization problems \cite{RN311}. This application was used to orchestrate and train a distributed prediction models using the proposed machine learning architecture. 

\subsection{Datasets}
For the experiments described above, we used the multiple public IoT datasets in our simulations. A comprehensive overview of the structure of these datasets is given below. 

\noindent \textbf{UNSW-NB15\footnote{https://www.unsw.adfa.edu.au/unsw-canberra-cyber/cybersecurity/ADFA-NB15-Datasets/}:} This dataset consists of transaction data (each containing 49 numerical and categorical features, i.e. $\in \mathbb{R}^{49}$) traced from a simulated intrusion detection system (IDS). The aforementioned IDS captures intrusive actions by malicious sensor providers similar to what was introduced in Section \ref{introduction}. Each record in the dataset corresponds to a transaction indicating either a benign behaviour and or one of nine types of attacks scenarios. We labelled each sample as \textit{benign} or \textit{harmful} based on whether they correspond to a benign or attack scenario. This dataset was first normalized and then divided into 100 randomly-sized (n$\left. \in \right.$[200, 2000]) smaller datasets forming an aggregate of 110892 examples. Random noise was also added to each dataset via flipping the labels of randomly picked samples to mimic a Non-IID dataset. The resulting datasets (with a training-to-test split ratio of 70:30) were used to train the distributed prediction models for each simulated MEC-environment.

\noindent \textbf{Gas Sensor Data\footnote{https://dataverse.harvard.edu/}:} This dataset contains 57982 training samples ($\in \mathbb{R}^{12}$) collected from eight gas sensors, a humidity and a temperature sensor, which were part of a home activity monitoring system. Each training example is labelled with one of two classes based on the type of chemical signal being monitored (i.e. \textit{banana} or \textit{wine}). To simulate an MEC-based IoT environment as well as the non-IID behavior of data accumulated in them, we formed 100 randomly-sized (n$\left. \in \right.$[100, 2000]) splits of the original dataset, 
and added random noise. Each resulting dataset was then divided into training and test sets with a training-to-test split ratio of 70:30.

\noindent\textbf{Human Activity Recognition\footnote{https://archive.ics.uci.edu/ml/machine-learning-databases/00366/}:} This dataset contains 20827 sensed human activity data samples gathered through an activity recognition system designed for smart homes. The training samples ($\in \mathbb{R}^{6}$) are collected from multiple sensors, and labelled under seven main activity categories (i.e. walking, standing, sitting, cycling, two modes of bending and lying). We re-labelled them to predict between \textit{walking} and \textit{other} activities, and created 100 randomly-sized (n$\left. \in \right.$[100, 300]) splits for the experiments with a training-to-test ratio of 70:30.

\section{Results and Discussion}
\label{results-and-discussion}

Results of our experiments showed that the binary SVM classifiers trained by NL framework consistently outperformed the Mocha FMTL framework solving the clustered-regularized multi-task model used by \cite{RN319} in terms of accuracy on all three datasets (TABLE-\ref{table-average-prediction-accuracy}). The aforementioned observations align quite well with multiple aspects we discussed in Section-\ref{algorithm-comparison}. For instance, it seems reasonable to assume that the \textit{exactness} of the optimization steps associated with NL mathematical framework led to the higher accuracy over Mocha FMTL framework. Furthermore, Within its optimization framework, NL uses a full distributed batch gradient step at every iteration. The stability of such a gradient update can also be a contributing factor behind NL framework's higher performance over Mocha FMTL framework in terms of prediction accuracy.

In addition to that, NL framework also performed significantly well in comparison to the simulated MEC-local binary SVM classifiers as well as the global SVM classifier. On the other hand, Mocha FMTL framework too showed improvements (albeit small) against the MEC-local SVM classifiers, which can be attributed to its ability to allow knowledge sharing among tasks. However, the performance of Mocha FMTL framework was considerably lower than that of the global SVM classifier. This observation contradicts with the results reported in the first work that introduced Mocha FMTL framework \cite{RN319}. We believe, the low-skew in the datasets used in our experiments would have favoured the global SVM classifier in this particular context.

\begin{table}[h!]
\centering
\begin{tabular}{cccc} \toprule
    Model & UNSW-NB15 & Gas Sensor & Human Activity\\ \midrule
    Global  & 99.96  & 56.53 & 75.32\\\midrule
    Local  & 82.17 & 50.23 & 73.69\\\midrule
    Mocha  & 83  & 50.85 & 73.93\\\midrule
    Network Lasso  & 100  & 61.89 & 76.7\\\bottomrule
\end{tabular}
\caption{Average prediction accuracy (\%) of the evaluated models.}
\label{table-average-prediction-accuracy}
\end{table}


\begin{figure*}
  \includegraphics[width=1\linewidth, height=3.8cm]{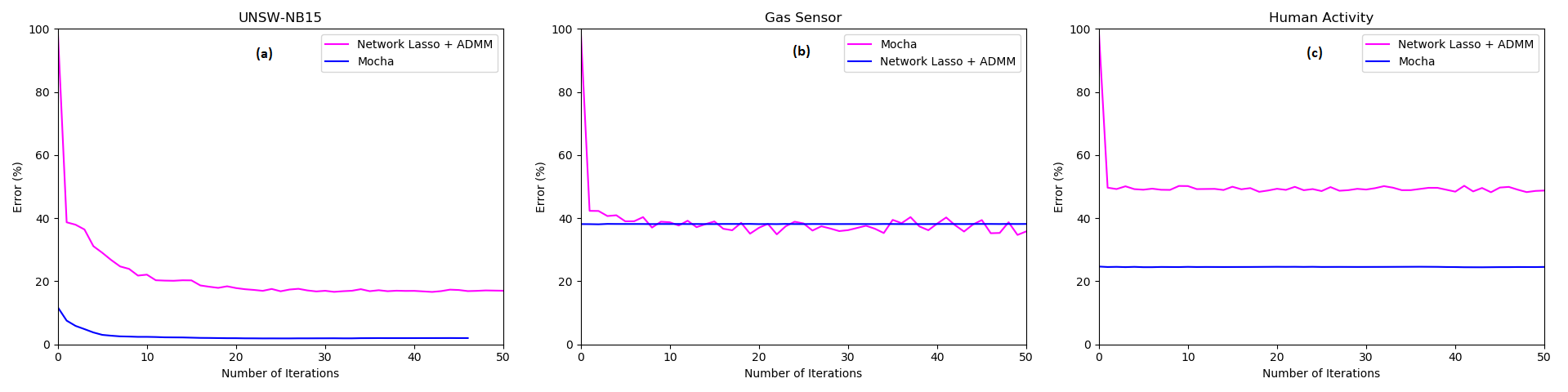}
  \caption{Rate of convergence against the number of communication iterations for all datasets evaluated.}\label{fig:convergence}
\end{figure*}

Meanwhile, the experiments we conducted on testing the communication efficiency of the two approaches revealed that the number of iterations needed by NL framework to reach an arbitrary $\epsilon$-accuracy is much less than that of Mocha FMTL framework atop both UNSW-NB15 and Human Activity Recognition datasets. The remaining dataset, however, showed almost comparable characteristics for both NL and Mocha FMTL approaches (Fig-\ref{fig:convergence}). As per the Fig-\ref{fig:convergence}, NL exhibited better accuracy than its counterpart even within on the intermediate iterations of the algorithm, well before reaching its desired convergence criteria. 

NL carries two key parameters in the form of $\lambda$, which enforces clustering among the prediction models trained by MEC environments, and $\rho$, which forces a significant impact on the convergence of the underlying optimization framework \cite{RN192}. In our experiments, we inferred a suitable value $\lambda$ that reduced the cross-validation error and used $\rho = 1.0$ as a suitable default value. However, these parameters being data-dependent (and therefore, are application-specific), inferring appropriate values for them through hyperparameter tuning strategies such as cross validation is challenging due to the distributed nature in which data is accumulated. Therefore, a more pragmatic approach that fits into the characteristics of MEC-based systems would be to run NL framework on its regularization path as described in \cite{RN192}, which comes at a cost of a lesser communication efficiency. In other words, starting from $\lambda$=0 or other suitable initial value (as recommended in \cite{RN192}), we can run NL framework once per given value of $\lambda$ while monotonically increasing its value. Once Algorithm-2 converges for a given value of $\lambda$, it is then restarted with $\lambda$ suitably incremented. The hope is that, with such a strategy, we attempt a \textit{warm-start} (i.e. using the the optimal solution achieved in the previous run of the algorithm as the initial point for the next run.) towards an optimal solution at each incremented value of $\lambda$, which can reduce the number of iterations needed for the algorithm to converge. In contrast, running Mocha FMTL in a distributed setting tends to be quite straight-forward primarily due to the fact that it has fewer hyper-parameters that need tuning.

Furthermore, an emerging class of machine learning applications in the context of MEC-based IoT systems is using non-convex deep learning models to allow predictive analytics on many real-world applications \cite{RN333, RN324, RN330}. Even though both Mocha FMTL and NL parallized by ADMM frameworks do not support such non-convex deep learning models out-of-the-box, there can be an opportunity to support convexified variants of them, with suitable trivial extensions to the underlying algorithmic framework \cite{RN319}. However, in the context of regularized loss minimization, the \textit{network lasso penalty} ($\norm{w_j - w_k}_2$ in problem \eqref{network-lasso}) is likely to generate better results predominantly upon robust and stable updates to model parameters $(w_j, w_k)$ at each iteration of the Algorithm-2. In the face of highly stochastic models trained in a distributed manner where model parameters shared by each distributed model at every iteration tend to be \textit{unstable}, there can be a possibility that such a regularizer might slow-down the convergence of the underlying algorithmic framework. 

In conclusion, it is evident from the empirical results above that the better machine learning framework between Mocha FMTL and NL frameworks in the context of MEC-based IoT environments depends on multiple factors. However, the obtained results suggest that NL outperforms Mocha FMTL framework not only in terms of prediction performance, but also the number of communication iterations needed in order to reach a given accuracy bound. Mocha FMTL framework, meanwhile, promises vital features such as fault tolerance, straggler avoidance and minimal need of parameter tuning, in the core of its algorithmic framework. This greatly enhances its adaptability to dynamic environments such as MEC-based IoT environments. Having considered the aforementioned aspects, in a more general context, it can be concluded that the NL framework is a better choice to tackle the challenges in MEC-based IoT environments.

\section{Conclusions and Future Work}
\label{conclusion}
We analysed the strengths and weaknesses of FMTL and NL paradigms and comprehensively compared their performances in MEC-based IoT environments. The results demonstrated that, even though Mocha FMTL framework has a collection of promising features, NL is a better fit in the MEC-based IoT environment due to its higher prediction performance and comparatively lower number of communication iterations needed to achieve the expected accuracy. In the future, we plan to improve NL in terms of the used machine learning model and task grouping based on our previous works \cite{qin2005}\cite{gong2016}.

\bibliographystyle{IEEEtran}
\bibliography{ijcnn-2020}

\begin{thebibliography}{10}
\providecommand{\url}[1]{#1}
\csname url@samestyle\endcsname
\providecommand{\newblock}{\relax}
\providecommand{\bibinfo}[2]{#2}
\providecommand{\BIBentrySTDinterwordspacing}{\spaceskip=0pt\relax}
\providecommand{\BIBentryALTinterwordstretchfactor}{4}
\providecommand{\BIBentryALTinterwordspacing}{\spaceskip=\fontdimen2\font plus
\BIBentryALTinterwordstretchfactor\fontdimen3\font minus
  \fontdimen4\font\relax}
\providecommand{\BIBforeignlanguage}[2]{{%
\expandafter\ifx\csname l@#1\endcsname\relax
\typeout{** WARNING: IEEEtran.bst: No hyphenation pattern has been}%
\typeout{** loaded for the language `#1'. Using the pattern for}%
\typeout{** the default language instead.}%
\else
\language=\csname l@#1\endcsname
\fi
#2}}
\providecommand{\BIBdecl}{\relax}
\BIBdecl

\bibitem{RN23}
M.~T. Beck, M.~Werner, S.~Feld, and S.~Schimper, ``Mobile edge computing: A
  taxonomy,'' in \emph{AFIN}.\hskip 1em plus 0.5em minus 0.4em\relax Citeseer,
  2014, Conference Proceedings, pp. 48--55.

\bibitem{RN323}
P.~Abeysekara, H.~Dong, and A.~Qin, ``Machine learning-driven trust prediction
  for mec-based iot services,'' in \emph{2019 IEEE ICWS}.\hskip 1em plus 0.5em
  minus 0.4em\relax IEEE, 2019, Conference Proceedings, pp. 188--192.

\bibitem{RN319}
V.~Smith, C.-K. Chiang, M.~Sanjabi, and A.~S. Talwalkar, ``Federated multi-task
  learning,'' in \emph{Advances in Neural Information Processing Systems},
  2017, Conference Proceedings, pp. 4424--4434.

\bibitem{RN192}
D.~Hallac, J.~Leskovec, and S.~Boyd, ``Network lasso: Clustering and
  optimization in large graphs,'' in \emph{SIGKDD}.\hskip 1em plus 0.5em minus
  0.4em\relax ACM, 2015, Conference Proceedings, pp. 387--396.

\bibitem{RN326}
Y.~Zhang, Y.~Wei, and Q.~Yang, ``Learning to multitask,'' in \emph{NIPS}, 2018,
  Conference Proceedings, pp. 5771--5782.

\bibitem{RN325}
Y.~Zhang and Q.~Yang, ``A survey on multi-task learning,'' \emph{arXiv preprint
  arXiv:1707.08114}, 2017.

\bibitem{RN211}
S.~Boyd, N.~Parikh, E.~Chu, B.~Peleato, and J.~Eckstein, ``Distributed
  optimization and statistical learning via the alternating direction method of
  multipliers,'' \emph{Foundations and Trends in Machine learning}, vol.~3,
  no.~1, pp. 1--122, 2011.

\bibitem{RN328}
M.~T. Beck and M.~Maier, ``Mobile edge computing: Challenges for future virtual
  network embedding algorithms,'' in \emph{ADVCOMP}, vol.~1, 2014, Conference
  Proceedings, p.~3.

\bibitem{RN329}
T.~X. Tran, A.~Hajisami, P.~Pandey, and D.~Pompili, ``Collaborative mobile edge
  computing in 5g networks: New paradigms, scenarios, and challenges,''
  \emph{IEEE Communications Magazine}, vol.~55, no.~4, pp. 54--61, 2017.

\bibitem{RN335}
R.~Zhang and J.~Kwok, ``Asynchronous distributed admm for consensus
  optimization,'' in \emph{ICML}, 2014, Conference Proceedings, pp. 1701--1709.

\bibitem{RN311}
S.~Diamond and S.~Boyd, ``Cvxpy: A python-embedded modeling language for convex
  optimization,'' \emph{The Journal of Machine Learning Research}, vol.~17,
  no.~1, pp. 2909--2913, 2016.

\bibitem{RN337}
S.~Shalev-Shwartz and T.~Zhang, ``Stochastic dual coordinate ascent methods for
  regularized loss minimization,'' \emph{Journal of Machine Learning Research},
  vol.~14, no. Feb, pp. 567--599, 2013.

\bibitem{RN270}
W.~Zhong and J.~Kwok, ``Fast stochastic alternating direction method of
  multipliers,'' in \emph{ICML}, 2014, Conference Proceedings, pp. 46--54.

\bibitem{RN298}
H.~Ouyang, N.~He, L.~Tran, and A.~Gray, ``Stochastic alternating direction
  method of multipliers,'' in \emph{International Conference on Machine
  Learning}, 2013, Conference Proceedings, pp. 80--88.

\bibitem{RN226}
Y.~C. Hu, M.~Patel, D.~Sabella, N.~Sprecher, and V.~Young, ``Mobile edge
  computing—a key technology towards 5g,'' \emph{ETSI white paper}, vol.~11,
  no.~11, pp. 1--16, 2015.

\bibitem{RN333}
J.~Hochstetler, R.~Padidela, Q.~Chen, Q.~Yang, and S.~Fu, ``Embedded deep
  learning for vehicular edge computing,'' in \emph{IEEE/ACM SEC}.\hskip 1em
  plus 0.5em minus 0.4em\relax IEEE, 2018, Conference Proceedings, pp.
  341--343.

\bibitem{RN324}
H.~Li, K.~Ota, and M.~Dong, ``Learning iot in edge: Deep learning for the
  internet of things with edge computing,'' \emph{IEEE network}, vol.~32,
  no.~1, pp. 96--101, 2018.

\bibitem{RN330}
Z.~Zhou, H.~Liao, B.~Gu, K.~M.~S. Huq, S.~Mumtaz, and J.~Rodriguez, ``Robust
  mobile crowd sensing: When deep learning meets edge computing,'' \emph{IEEE
  Network}, vol.~32, no.~4, pp. 54--60, 2018.

\bibitem{qin2005}
A.~K. Qin and P.~N. Suganthan, ``Initialization insensitive{LVQ} algorithm
  based on cost-function adaptation,'' \emph{Pattern Recognition}, vol.~38,
  no.~5, pp. 773--776, 2005.

\bibitem{gong2016}
M.~Gong, Y.~Wu, Q.~Cai, W.~Ma, A.~K. Qin, Z.~Wang, and L.~Jiao, ``Discrete
  particle swarm optimization for high-order graph matching,''
  \emph{Information Sciences}, vol. 328, pp. 158--171, 2016.

\end{thebibliography}
\end{document}